\documentclass{ws-p10x7}
\def\alps{\alpha_s}
\def\ov{\overline}
\def\Del{\Delta}
\def\ddelg{\Delta\ov{\delta}_G}
\def\dsz{\Del S_Z}
\def\dtz{\Del T_Z}
\def\dr{\Del R}
\def\dof{\rm{d.o.f.}}
\def\wt{\widetilde}
\def\mzsq{m_Z^2}
\def\dgbl{\Delta g_L^b}
\def\dmw{\Delta m_W}
\def\disp{\displaystyle}
\def\gsim{{\mathop >\limits_\sim}}
\def\ie{{\it i.e.}}
\def\gev{{\rm GeV} }
\def\tev{{\rm TeV} }
\def\hph{\hphantom{-}}
\def\xa{x_\alpha}
\def\mw{m_W^{}}
\def\mt{m_t^{}}
\def\mh{m_{H_{\rm SM}}^{}}
\def\mz{m_Z^{}}
\def\ds{\Del S}
\def\dt{\Del T}
\def\du{\Del U}
\newcommand{\beq}{\begin{equation}}
\newcommand{\eeq}{\end{equation}}
\newcommand{\bea}{\begin{eqnarray}}
\newcommand{\eea}{\end{eqnarray}}
\def\PRD#1#2#3{Phys. Rev. {\bf D#1} (19#2) #3}
\def\PRL#1#2#3{Phys. Rev. Lett. {\bf #1} (19#2) #3}
\def\npb#1#2#3{Nucl. Phys. {\bf B#1} (20#2) #3}

\begin{document}
\title{Supersymmetry versus precision experiments revisited}
\author{Gi-Chol Cho}
\address{Department of Physics, Ochanomizu University,\\
	Bunkyo, Tokyo 112-8610, Japan\\
	E-mail: cho@phys.ocha.ac.jp}
\twocolumn[\maketitle\abstract{
We study constraints on the Minimal Supersymmetric Standard 
Model from electroweak experiments. 
We find that the light sfermions always make the fit worse 
than the Standard Model, while the light chargino generally 
make the fit slightly better through the oblique corrections. 
The best overall fit to the precision measurements are found 
when the mass of lighter chargino is about 100 GeV and the 
SU(2)$_L$ doublet sfermions are all much heavier. 
We find the slight improvement of the fit over the SM, where 
the total $\chi^2$ of the fit decreases by about one unit. 
}]
\section{Electroweak observables in the MSSM}
Precision measurements of the electroweak observables on 
the $Z$-pole at LEP1 and SLC~\cite{lepewwg98}, and 
the $W$-boson 
mass at LEP2 and Tevatron~\cite{wboson_moriond} are expected 
to give the stringent constraints on the Minimal Supersymmetric 
Standard Model (MSSM). 
The supersymmetric (SUSY) contributions to the electroweak 
observables are given through the oblique corrections 
and the process specific vertex/box corrections. 
It has been shown that the $Z$-pole observables are 
conveniently parametrized in terms of two oblique parameters 
$S_Z$ and $T_Z$~\cite{CH}, and the $Zf_\alpha f_\alpha$ 
vertex corrections, where $f$ denotes the fermion species 
and $\alpha$ is their chirality. 
The oblique parameters $S_Z$ and $T_Z$ are related to the 
$S$ and $T$ parameters~\cite{stu90}, 
\bea
S_Z &\equiv& S + R - 0.064 \xa, \\
T_Z &\equiv& T + 1.49R - \frac{\ddelg}{\alpha}, 
\label{eq:oblique}
\eea
where $\xa =\frac{1/\alpha(\mzsq)-128.90}{0.09}$ and 
$\ddelg/\alpha$ parametrize the hadronic uncertainty of 
the QED coupling and the corrections to the $\mu$-decay 
constant, respectively. 
The parameter $R$ is introduced as the difference of the 
$Z$-boson propagator corrections between $q^2=\mzsq$ and 
$q^2=0$. 
For convenience of later analysis, we introduce $\dsz$ and 
$\dtz$ as the shifts from $S_Z$ and $T_Z$ at the SM reference 
point, $\mt=175~\gev,\mh=100~\gev,\alps(\mz)=0.118$ and 
$1/\alpha(\mzsq)=128.90$, 
\bea
\dsz &=& \ds + \dr - 0.064 \xa, 
\\
\dtz &=& \dt + 1.49 \dr - \frac{\ddelg}{\alpha}. 
\eea
In addition to $\dsz$ and $\dtz$, we adopt the $W$-boson mass 
$\mw(\gev)=80.402 + \dmw$ as the third oblique parameter instead 
of the $U$-parameter~\cite{stu90}:
\bea
\dmw &=& -0.288 \ds + 0.418 \dt + 0.337 \du \nonumber \\
	&&~~~+ 0.012\xa - 0.126\frac{\ddelg}{\alpha}. 
\eea
So the new physics contributions to the electroweak observables 
can be summarized by three oblique corrections $\dsz,\dtz$ and $\dmw$, 
and the non-oblique corrections $\Del g^f_\alpha$ and $\ddelg$. 
\section{Quantum corrections in the MSSM}
\begin{figure}
\epsfxsize200pt
\figurebox{120pt}{160pt}{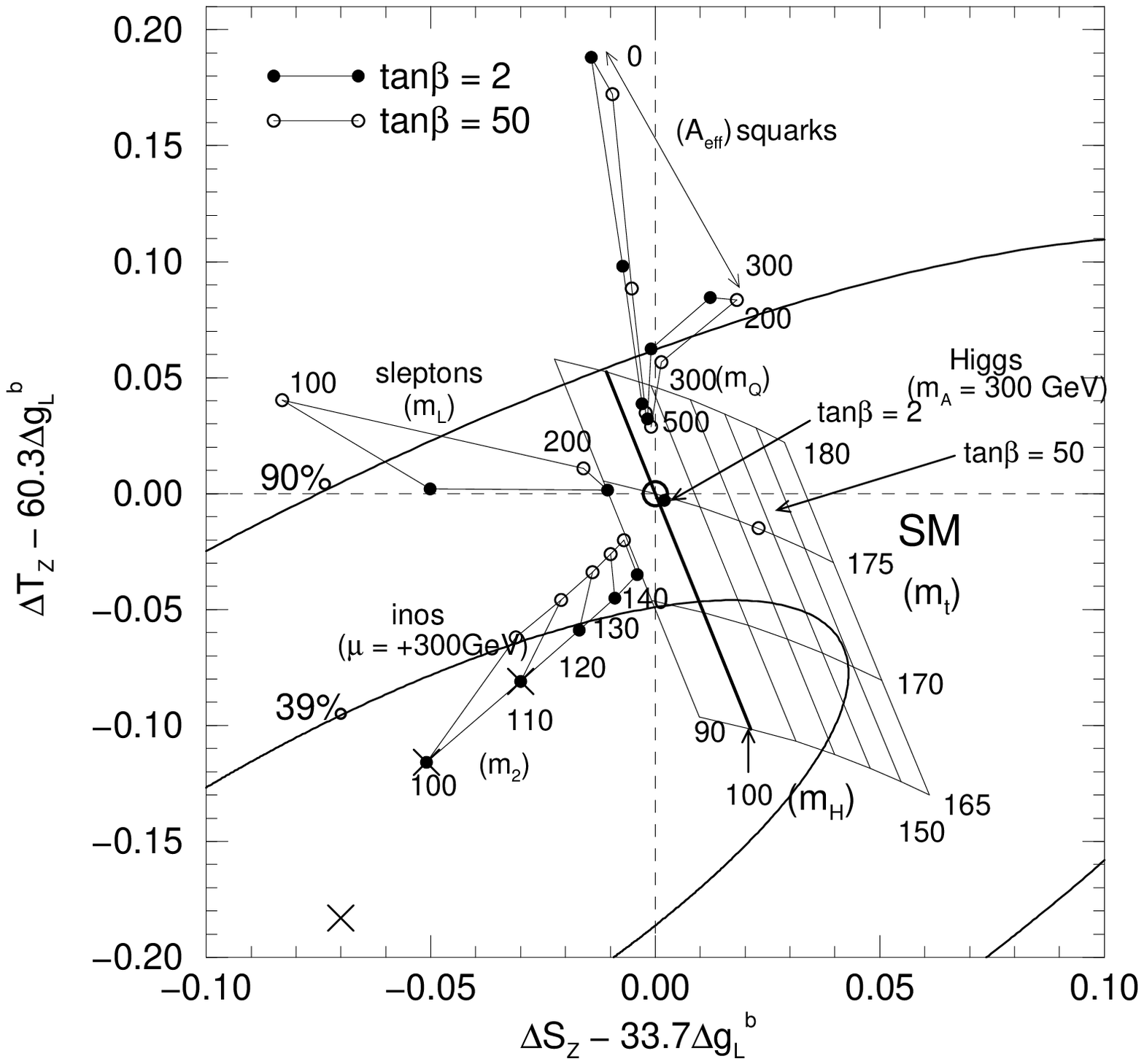}
\caption{
Supersymmetric contributions to $\dsz-33.7\dgbl$ and 
$\dtz-60.3\dgbl$. The symbol ($\times$) denotes the 
best fit from the electroweak data. 
The 39\% ($\Del \chi^2=1$) and 90\% ($\Del \chi^2=4.61$) 
contours are shown. The SM predictions are given for 
$\mt=165\sim 180~\gev$ and $\mh = 90\sim 150~\gev$. 
}
\label{fig:oblique}
\end{figure}

First we show the constraints on the oblique parameters from 
the experiments and study the SUSY contributions to them. 
Taking account of $\Del g_L^b$ which may have the non-trivial 
$\mt$-dependence even in the SM, we perform 
the 5-parameter fit ($\dsz,\dtz,\dmw,\dgbl,\alps(\mz)$) 
and find the following constraint: 
\bea
\begin{array}{l}
	\left. 
	\begin{array}{lcl}
	\Del S_Z -33.7 \Del g_L^b &=& -0.070 \pm 0.113 \\
	\Del T_Z -60.3 \Del g_L^b&=& -0.183 \pm 0.137 
	\end{array} \right \}
\rho = 0.89,
\\
\disp{ \Del m_W^{} = \hph 0.008 \pm 0.046 },\\
\disp{
\chi^2_{\sf min} = 15.4 + \Biggl( \frac{\Del g_L^b + 0.00086}{0.00077}
	\Biggr)^2,
}
\end{array}
\eea
where $\dof = 19-5 = 14$. 
We show the supersymmetric contributions 
to $\dsz-33.7\dgbl$ and $\dtz-60.3\dgbl$ individually for 
$\tan\beta=2$ and 50 in Fig.~\ref{fig:oblique}. 
The complete oblique corrections in the MSSM are given 
by their sum.  

The squark and slepton contributions are shown 
as functions of the soft SUSY breaking mass parameters 
$m_{\wt{Q}}$ and $m_{\wt{L}}$, respectively, by assuming 
that their universality between the left- and 
the right-handed components. 
The effects of the left-right mixing of the sfermions 
are examined by introducing the effective $A$-parameter, 
$A_{\rm eff} \equiv A_{\rm eff}^t = A_{\rm eff}^b$. 
For example, in the case of the stop, $A_{\rm eff}$ is given 
by $A_{\rm eff} = A_t - \mu\cot\beta$. 
It can be seen from the figure that the squark contribution 
makes the fit worse, because it always makes $\dtz$ larger 
than the SM. 
The presence of the left-right mixing (\ie~non-zero $A_{\rm eff}$) 
make the squark contribution to the $\dtz$-parameter slightly mild. 
The slepton contribution also makes the fit worse through 
the negative contribution to $\dsz$. 
We find that the contributions from the left-handed sfermions 
to $\dsz$ and $\dtz$ are much larger than those from 
the right-handed sfermions. 

We show the MSSM Higgs boson contributions to the oblique 
parameters for the $CP$-odd Higgs scalar mass $m_A=300~\gev$ 
and the SUSY breaking mass parameter $m_{\rm SUSY}=1~\tev$, 
which appears in the effective Higgs potential.   
The lightest Higgs boson mass $m_h$ in the figure is 
$106~\gev$ for $\tan\beta=2$, and $129~\gev$ for $\tan\beta=50$. 
We can see that the MSSM Higgs boson contributions to 
$(\dsz,\dtz)$ behave like the SM Higgs boson for $m_A \gsim 300~\gev$. 

We show the chargino/neutralino contributions to ($\dsz,\dtz$) 
as a function of the wino mass $M_2$ and for the higgsino mass 
$\mu=300~\gev$. 
The points with the $(\times)$ symbol in the figure are 
excluded from the direct search limit of the chargino mass 
($\sim 90~\gev$) at the LEP2 experiments. 
The chargino/neutralino contributions show the negative 
$\dsz$ and $\dtz$, which may improve the fit over the SM. 
This is essentially because of $\dr$ which resides both 
in $\dsz$ and $\dtz$.  
We find that the chargino contribution to $\dr$ is negative 
and large as compared to the sfermion contributions. 
For example, the contribution of the wino-like chargino 
to $\dr$ has the singularity 
$1/\sqrt{4M_2^2/\mzsq -1}$ when $M_2$ is close to a half 
of $\mz$. 
On the other hand, when $4M_2^2/\mzsq \gg 1$, 
$\dr$ is suppressed as $\mzsq/M_2^2$, but the coefficient 
of the wino contribution is found to be about 90 times 
larger than that of the right-handed slepton contribution. 
This large negative contribution to $\dr$ makes both 
$\dsz$ and $\dtz$ significantly negative when a relatively 
light chargino exists. 

\begin{figure}[t]
\epsfxsize200pt
\figurebox{120pt}{160pt}{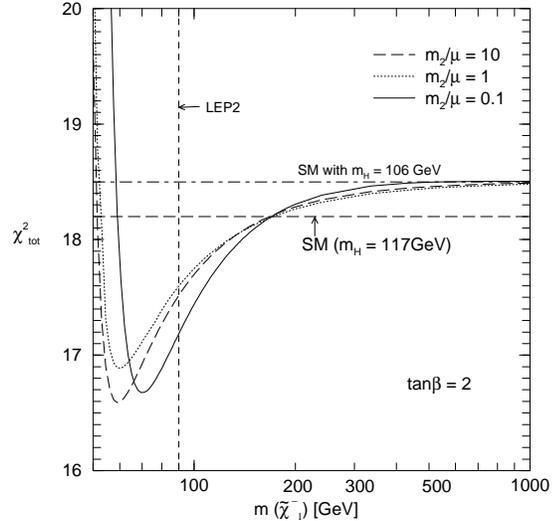}
\caption{ 
The total $\chi^2$ in the MSSM as a function of the lighter 
chargino mass $m_{\wt{\chi}^-_1}$ for $\tan\beta=2$. 
The SM best fit ($\chi^2 = 18.2$) is shown by the dashed 
horizontal line. 
The dot-dashed horizontal line shows the SM fit using 
$\mh = 106~\gev$ which is the lightest Higgs boson mass predicted 
in the MSSM.  
Three different $M_2$-$\mu$ ratio (10, 1, 0.1) are studied. 
The bound on $m_{\wt{\chi}^-_1}$ from the LEP2 experiment is 
shown by the dashed vertical line. 
}
\label{fig:global}
\end{figure}

Beside the oblique corrections, we have studied the non-oblique 
corrections such as the $Zff$ vertex corrections and/or 
the vertex/box corrections to the $\mu$-decay process 
in detail~\cite{CH}, and we found no improvement of the fit 
over the SM through the non-oblique corrections. 
Then, the best fit of the MSSM may be found when all sfermions 
and Higgs bosons are heavy enough, while the chargino is 
relatively light so that the radiative corrections 
in the MSSM are dominated by the chargino contribution to 
the oblique parameters. 
We perform the global fit to all electroweak data in the MSSM 
by assuming that all sfermions and heavy Higgs bosons masses 
are $1\tev$. 
Under this assumption, we show the total $\chi^2$ in the MSSM 
as a function of the lighter chargino mass $m_{\wt{\chi}^-_1}$ 
for $\tan\beta=2$ and $M_2/\mu=0.1, 1$ and 10 
in Fig.~\ref{fig:global}. 
The decoupling in the large SUSY mass limit is examined by 
comparing the MSSM fit with the SM fit at $\mh=m_h=106\gev$ 
rather than the SM best fit at $\mh=117\gev$. 
When the lighter chargino mass is around its lower mass bound 
from the LEP2 experiment, we find the slight improvement of 
the fit over the SM, where the total $\chi^2$ decreases by 
about one unit. 
The case of $\tan\beta=50$ shows the similar 
behavior~\cite{CH}. 
\section{Summary}

We have studied constraints on the MSSM from 
the electroweak precision experiments. 
Owing to the negative large contribution to $\dtz$ from 
the light chargino, the improvement of the fit over the SM 
is expected, if the left-handed sfermions are heavy enough 
to decouple from the electroweak processes. 
The global fit of the MSSM show that, if the masses of all 
sfermions and heavy Higgs bosons are $1~\tev$, the total 
$\chi^2$ in the MSSM decreases by about one unit comparing 
with the SM when the lighter chargino mass is close to its 
direct search limit from LEP2. 

\section*{Acknowledgments}
The author would like to thank Kaoru Hagiwara for fruitful 
collaboration which this report is based upon.

\end{document}